\pdfoutput=1
\documentclass[aps,prc,twocolumn,floatfix,superscriptaddress,nofootinbib,preprintnumbers,longbibliography,amsmath,amsthm,amssymb]{revtex4-2}
\usepackage{graphicx}

\usepackage{amsfonts}
\usepackage{color}
\usepackage[colorlinks=true,linkcolor=blue,citecolor=blue]{hyperref}
\usepackage{dcolumn}
\usepackage{bm}
\usepackage{braket}

\newcommand{\be}{\begin{equation}}
\newcommand{\ee}{\end{equation}}

\newcommand{\br}{\boldsymbol{r}}

\newcommand{\ii}{\mathrm{i}}

\begin{document}
\allowdisplaybreaks[1]

\title{Cranked Skyrme--Hartree--Fock--Bogoliubov approach for a mean-field description of nuclear rotations near the drip line
}

\author{Kenichi Yoshida}
\email[E-mail: ]{kyoshida@ruby.scphys.kyoto-u.ac.jp}
\affiliation{Department of Physics, Kyoto University, Kyoto, 606-8502, Japan}

\preprint{KUNS-2893}
\date{\today}

\begin{abstract}
To describe the yrast states in weakly bound nuclei, 
I directly solve the coordinate-space cranked Skyrme--Hartree--Fock--Bogoliubov equation 
on a three-dimensional lattice with the continuum states discretized in a box. 
After the numerical demonstration for the ground-state band in a medium-mass nucleus, 
I apply the newly-developed method to neutron-rich even-$N$ Mg isotopes. 
I find that the appearance of the significantly low $I^\pi=2^+$ state in $^{40}$Mg 
is mainly due to the suppression of pairing.  
The calculation predicts that 
the $2^+$ state in $^{42}$Mg appears as high in energy as in $^{34\text{--}38}$Mg whereas 
the triaxial deformation is enhanced in non-zero spin states. 
The present numerical framework  
offers a practical approach for investigating the near yrast states systematically 
and revealing structures unique in drip-line nuclei.   

\end{abstract}

\maketitle

\section{Introduction}

A diverse variety of modes of motion show up in nuclei. 
To explore unique structures in exotic nuclei, 
various spectroscopic studies have been carried out via decays and making use of direct reactions. 
Nuclear deformation is a collective phenomenon that has long attracted interest; 
a nucleus is deformed as it moves away from the magic number due to the many-body correlation. 
It has been turned out that the shell structure evolves as the conventional magic numbers disappear 
and new ones appear instead, depending on the combination of the particle numbers of protons and neutrons~\cite{ots20}. 
Energies of excited nuclear states are often among the first quantities accessible in experiments 
and have been used as an indicator of the changing shell structure and the onset of deformation~\cite{gad15}.

Rotational motion is a manifestation of the spontaneous breaking of the rotational symmetry~\cite{BM2}. 
The cranking approximation provides an intuitive picture of the rotation of quantum systems, 
and the cranked shell model is a standard tool to investigate the microscopic structure of nuclear rotations 
near the ground state and at high spins~\cite{voi83,ben86,gar86,sat04}. 
The cranked shell model in the framework of the nuclear energy-density functional (EDF)---the cranked Kohn--Sham (KS) approach---has 
provided a systematic and quantitative description of the (near) yrast states from light to heavy nuclei~\cite{ben03,vre05,rob18}. 
The accumulation of experimental data for low spins in neutron-rich nuclei 
has stimulated 
the application of the cranked KS approach 
to the exploration of the rotational motions unique in weakly bound nuclei.

Since the moments of inertia of neutron-rich nuclei are sensitive to the pairing and weak binding~\cite{yam08}, 
one needs to solve the cranked KS--Bogoliubov (KSB) equation in the coordinate space
to describe properly the weak binding and the pairing embedded in the continuum~\cite{dob84}, 
which is computationally demanding because many symmetries are broken and the 
quasiparticle wave functions are spatially extended. 
The low-lying collective states have been investigated microscopically in beyond-mean-field approaches, 
such as the projected shell model~\cite{har95}, the generator-coordinate method and the collective Hamiltonian~\cite{ben03,vre05,rob18,nak16}. 
Including the continuum effects in these approaches is much more demanding.

The island of inversion has been the subject of much experimental and theoretical interest~\cite{sor08}. 
A systematic calculation for the Mg isotopes in the mean-field approximation 
produces a spherical configuration in $^{32}$Mg, a soft potential energy surface in $^{34}$Mg, 
and a prolate configuration in $^{36,38,40}$Mg for the ground state~\cite{ter97}. 
It has been clarified that 
the shape fluctuation and the correlation beyond the mean-field approximation 
is significant in $^{32}$Mg~\cite{kim02,rod02,hin11}. 
Experimentally, 
the measurement of not only the low excitation energy of the first $I^\pi=2^+$ state 
but the energy of the $4^+$ state and their ratio with $R_{4/2}$ being greater than three has 
revealed a well-deformed structure in $^{34,36,38}$Mg~\cite{doo13}. 
A significantly low $2^+$ state in $^{40}$Mg 
is not reproduced by any theoretical models~\footnote{The configuration-mixing calculation using the Gogny force 
taking only the axial symmetry produces the increase in $E(2^+_1)$ from $N=22$ to $26$ contrary to the measurements and a drop in energy at $N=28$~\cite{rod02,shi16}. 
However, the inclusion of the triaxial deformation reproduces the isotopic dependence up to $^{38}$Mg, and the drop at $^{40}$Mg is washed out.~\cite{rod16}. 
The relativistic approach in Ref.~\cite{yao11} produces a strong neutron-number dependence.}, 
and it indicates 
a unique feature associated with the weak binding~\cite{cra19}. 
As a coherent contribution of the pairing in the continuum states 
induces an enhanced quadrupole transition to the low-lying vibrational state 
in the Mg isotopes~\cite{yos06a,yos06,yos08b,yos09a,yam19}, 
the roles of the weak binding and the continuum coupling 
in the rotational motions are interesting to study.

I investigate in this article the low-spin states in the neutron-rich Mg isotopes close to the drip line. 
Then, I try to clarify the roles of the weak binding and excess neutrons in the low-lying excited states.  
To this end, I develop a new framework of the cranked shell model within the nuclear EDF approach, 
which is capable of handling nuclides with arbitrary mass numbers~\cite{ben03,nak16}. 

\section{Method}

To describe the (near) yrast states with proper account of the pairing in the continuum states, 
I directly solve the coordinate-space cranked Skyrme-KSB 
or Hartree--Fock--Bogoliubov equation 
in the quasiparticle basis: 
\begin{align}
\sum_{\sigma^\prime}
\begin{bmatrix}
h^{q \prime}_{\sigma \sigma^\prime}(\br)
& \tilde{h}^q_{\sigma \sigma^\prime}(\br) \\
4\sigma \sigma^\prime \tilde{h}^{q*}_{-\sigma -\sigma^\prime}(\br) 
& -4\sigma \sigma^\prime h^{q\prime*}_{-\sigma -\sigma^\prime}(\br)
\end{bmatrix}
\begin{bmatrix}
\varphi^{q}_{1,\alpha}(\br \sigma^\prime) \\
\varphi^{q}_{2,\alpha}(\br \sigma^\prime)
\end{bmatrix} \notag \\
= E_{\alpha}
\begin{bmatrix}
\varphi^{q}_{1,\alpha}(\br \sigma) \\
\varphi^{q}_{2,\alpha}(\br \sigma)
\end{bmatrix}, \label{HFB_eq}
\end{align}
which is obtained by extending the formalism developed for describing the ground-state properties of 
even-even nuclei near the drip line~\cite{dob84}. 
Here 
the single-particle Routhian and the pair Hamiltonian 
are defined by using a Skyrme EDF combined with a pairing functional $E[\rho,\tilde{\rho},\tilde{\rho}^*]$ as 
$h^{q \prime}_{\sigma \sigma^\prime}(\br)
=\frac{\delta E[\rho,\tilde{\rho},\tilde{\rho}^*]}{\delta \rho^q_{\sigma^\prime \sigma}(\br)}-(\lambda^{q}+\omega_{\rm rot} j_z)\delta_{\sigma \sigma^\prime} $ and 
$\tilde{h}^q_{\sigma \sigma^\prime}(\br)=\frac{\delta E[\rho,\tilde{\rho},\tilde{\rho}^*]}{\delta \tilde{\rho}^{q*}_{\sigma^\prime \sigma}(\br)}$. 
I define the $z$-axis as a quantization axis of the intrinsic spin and consider the system rotating uniformly about the $z$-axis. 
I take the natural units: $\hbar =c=1$. 

The Skyrme-KSB equation in the three-dimensional (3D) Cartesian-mesh has been solved 
by employing the contour integral technique and the shifted Krylov subspace method for the 
Green's function~\cite{jin17,kas20} to circumvent the 
successive diagonalization of the matrix with huge dimension. 
Very recently, the direct diagonalization of the KSB Hamiltonian in the 3D Cartesian-mesh has been achieved~\cite{jin21}.
The numerical procedure to solve Eq.~(\ref{HFB_eq}) in the present study 
is basically the same as in solving the cranked KS equation in Refs.~\cite{sak20,yos21a}.
I impose the reflection symmetry about the $(x, y)$-, $(y,z)$- and $(z, x)$-planes to reduce the computational time. 
Thus, the parity $\mathfrak{p}_k$ $(=\pm 1)$ and $z$-signature $r_k$ ($=\pm \ii$) are a good quantum number. 
I solve Eq.~(\ref{HFB_eq}) by diagonalizing the KSB Hamiltonian directly 
in the 3D Cartesian-mesh representation with the box boundary condition. 
Thanks to the reflection symmetries, I have only to consider explicitly the octant region in space 
with $x\ge0$, $y\ge0$, and $z\ge0$; see Refs.~\cite{bon87,oga09} for details.
I use a 3D lattice mesh $x_i=ih-h/2, y_j=jh-h/2, z_k=kh-h/2 \ \  (i,j,k=1,2,\cdots M)$ with a mesh size $h$. 
The dimension of the KSB Hamiltonian is thus $8M^3$.
To check the convergence of the results with respect to the box size and to investigate the effect of the weak binding, 
I change $h$ and $M$ in the discussion below. 
The differential operators are represented by the use of the 9-point formula of the finite difference method.
For diagonalizing the matrix of Eq.~(\ref{HFB_eq}), I use the ScaLAPACK {\sc pdsyev} subroutine~\cite{ScaLAPACK}. 
A modified Broyden's method~\cite{bar08} is utilized to calculate new densities during the selfconsistent iteration. 
For each iteration, it took 13.0, 52.5, and 91.0 core-hours for $M=12$, 14, and 16, respectively at the Yukawa-21 computer facility.
The quasiparticle energy is cut off at 60 MeV, 
which has almost no significant change in computational time but determines the memory required during the calculation. 

\begin{figure}[t]
\begin{center}
\includegraphics[scale=0.5]{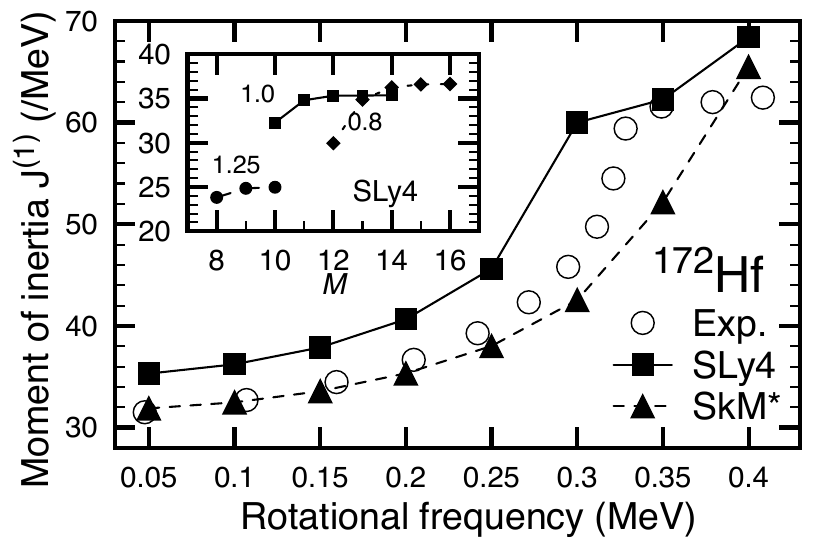}
\caption{\label{fig:172Hf} 
Kinematic moments of inertia $\mathcal{J}^{(1)}$ 
as functions of the rotational frequency. The experimental data~\cite{NNDC} are denoted by open symbols. 
Dependence of $\mathcal{J}^{(1)}$ evaluated at $\omega_{\rm rot}=0.05$ MeV 
on the parameters $h$ and $M$ is shown in the inset. 
}
\end{center}
\end{figure}

\section{Results and discussion}

\subsection{Ground-state band in $^{172}$Hf}

To see the validity of the present framework, I perform the calculation for the ground-state 
rotational band in $^{172}$Hf as a typical example of the collective rotation. 
Figure~\ref{fig:172Hf} shows the calculated kinematic moments of inertia $\mathcal{J}^{(1)}$ 
as functions of the rotational frequency. Here, $\mathcal{J}^{(1)}$ is defined by $J_z/\omega_{\rm rot}$.
I employed the SLy4~\cite{cha98} and SkM*~\cite{bar82} functionals together with 
the Yamagami--Shimizu--Nakatsukasa (YSN) pairing EDF in Ref.~\cite{yam09}. 
The inset shows the calculated $\mathcal{J}^{(1)}$ at $\omega_{\rm rot}=0.05$ MeV obtained by varying $h$ and $M$. 
One sees the results are converged at about $11\text{--}12$ fm with respect to the box size. 
The calculated rms radius is 5.37 fm and 5.28 fm for neutrons and protons, respectively. 
Thus, a rough estimate for a sufficient box size 
is that one needs a box about twice as large as the rms radius.
Since the mesh size $h=1.0$ fm gives a reasonable convergence as found in the early studies~\cite{dav80,bon85}, 
I use $M=12$ and $h=1.0$ fm in the following. 
It is noticed that a systematic numerical investigation in Ref.~\cite{taj01} revealed that 
the 3D mesh calculation gives a remarkably high precision with apparently coarse meshes such as $h=1.0$ fm.

The present model describes well the low spin states and the band crossing. 
Around $\omega_{\rm rot}=0.25$ MeV, the alignment of neutrons in the $i_{13/2}$ orbital occurs for the case of SLy4 
whereas this is lower than the measurement $\sim 0.3$ MeV. 
One sees that the rotational property beyond the band crossing is also reasonably described. 
The SkM* functional describes the alignment around $\omega_{\rm rot}=0.35$ MeV, 
however the level crossing is more gentle than in the case of SLy4. 

\begin{figure}[t]
\begin{center}
\includegraphics[scale=0.5]{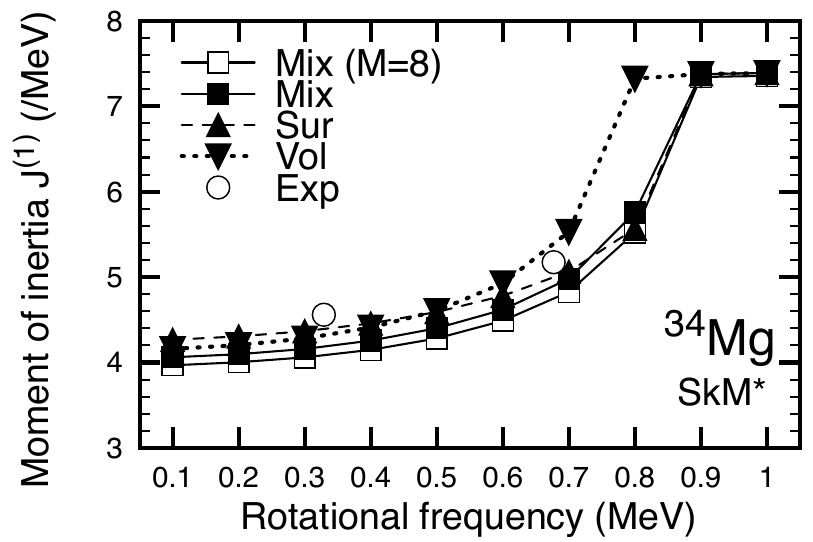}
\caption{\label{fig:34Mg} 
Similar to Fig.~\ref{fig:172Hf} but for $^{34}$Mg. 
The results obtained by using several pairing interactions are displayed. 
The experimental data are obtained from Ref.~\cite{mic14}.
}
\end{center}
\end{figure}

\subsection{Neutron-rich Mg isotopes}

Then, I investigate the low-spin yrast states in the neutron-rich Mg isotopes. 
I use the SkM* functional~\cite{bar82} and the so-called mixed-type pairing interaction 
with the strength $V_0=-295$ MeV fm$^3$ as in Refs.~\cite{yos09a,yos09b},
in which the low-frequency vibrations were investigated.   
The ground-state properties thus obtained are summarized in Table 1 of Ref.~\cite{yos09a}. 
I found that the triaxiality is negligibly small: $\gamma<1^\circ$ in low spins. 
The chemical potential is a key quantity governing the spatial structure of quasiparticle wave functions: 
$-4.17, -3.26, -2.42, -1.61$ MeV for neutrons in $^{34,36,38,40}$Mg, respectively. 
When the two-basis method, in which the KSB Hamiltonian is diagonalized in a truncated single-particle basis 
obtained by solving the KS equation in the Cartesian-mesh~\cite{ben03}, 
is employed for such weakly bound nuclei, 
the single-particle scattering states enter the pairing window. 
Therefore, 
the convergence with respect to the number of basis states has to be carefully examined 
as the densities are spatially localized for $\lambda <0$.
In the present case, however, the densities are always calculated to be localized 
because the full KSB Hamiltonian is directly diagonalized~\cite{dob84}. 

Figure~\ref{fig:34Mg} shows the calculated $\mathcal{J}^{(1)}$ of $^{34}$Mg, and compares 
with the experimental data~\cite{mic14}. 
The measured $R_{4/2}$ value is 3.06, which is lower than that of the rigid rotor and close to the calculated value 2.96.
The calculation reproduces well the slight increase in $\mathcal{J}^{(1)}$ due to the weakening of pairing. 
To see the effect of weak binding, the results obtained by using $M=8$ are also included in Fig.~\ref{fig:34Mg}. 
The difference between the cases with $M=8$ and 12 is not very significant, 
which indicates that the spatial extension of neutrons is not important. 
It is noted that 
the calculated rms radius is 3.49 fm and 3.14 fm for neutrons and protons, respectively. 
That I obtained the converged results with a box size of 7--8 fm 
is in accordance with the above example for $^{172}$Hf. 

The role of the density dependence of the pairing interaction has been discussed 
in the study of the superdeformed states~\cite{ter95}: the density dependence results in a retarded alignment. 
I investigate here the density dependence of the pairing interaction. 
To this end, I use the volume- and surface-type pairing interactions.  
I determined the strength to keep the calculated pairing gap as obtained with the mixed pairing at $\omega_{\rm rot} = 0$ MeV: 
$V_0=-211$ and $-423$ MeV fm$^3$ for the volume and surface pairing, respectively. 
The pairing gap is defined by $\Delta^q=\int d\br \tilde{h}^q(\br)\tilde{\rho}^{q*}(\br)/\int d\br \tilde{\rho}^{q*}(\br)$.
Notice that the protons are unpaired at $\omega_{\rm{rot}}=0$. 
In low $\omega_{\rm rot}$, 
the difference among three types of interaction is relatively small. 
However, the volume pairing 
gives a faster increase in $\mathcal{J}^{(1)}$ 
similarly to the finding in Ref.~\cite{ter95}.

\begin{figure}[t]
\begin{center}
\includegraphics[scale=0.5]{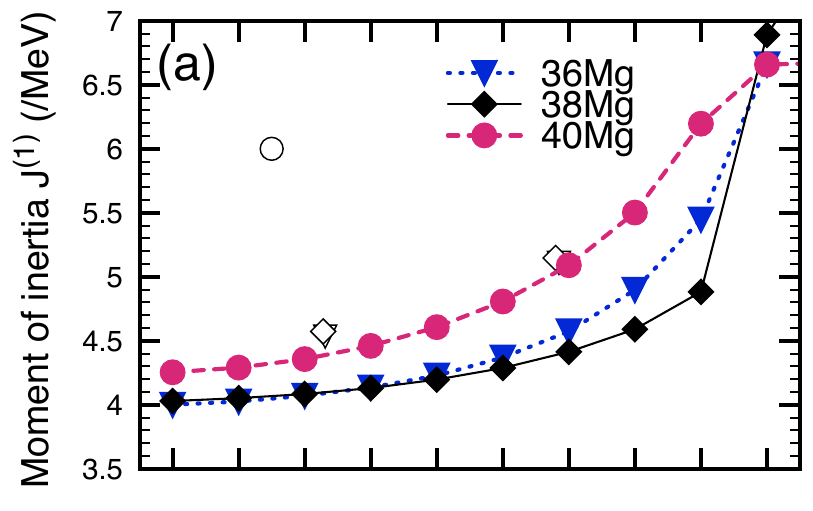}\\
\includegraphics[scale=0.5]{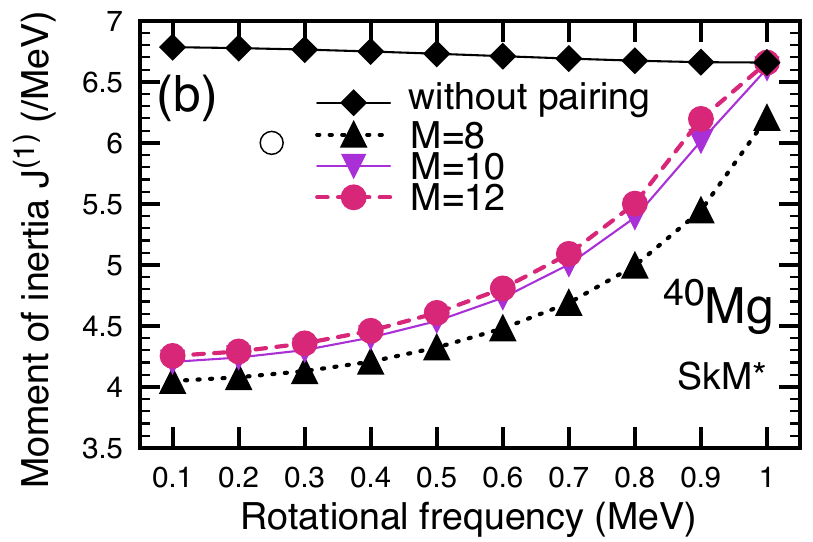}
\caption{\label{fig:Mg} 
(a) Similar to Fig.~\ref{fig:34Mg} but for $^{36,38,40}$Mg. 
The experimental data, denoted by open symbols, are taken from Refs.~\cite{doo13,cra19}.
(b) Calculated $\mathcal{J}^{(1)}$ for $^{40}$Mg obtained with the box size of $M=8,10$, and 12. 
Results without the pairing are also depicted.   
}
\end{center}
\end{figure}

Next, I investigate the rotational property of $^{36,38,40}$Mg located close to the drip line. 
Figure~\ref{fig:Mg}(a) displays the calculated $\mathcal{J}^{(1)}$ together with the experimental data~\cite{doo13,cra19}.
The calculated $\mathcal{J}^{(1)}$ for $^{36,38}$Mg in low $\omega_{\rm{rot}}$ is similar to 
the one for $^{34}$Mg, and smaller than that for $^{40}$Mg. 
This is consistent with the calculation in Ref.~\cite{yos09a}, 
where the moments of inertia were evaluated using the Thouless--Valatin procedure 
in the framework of the Skyrme EDF-based QRPA~\cite{yos08}. 
A higher value for $\mathcal{J}^{(1)}$ of $^{40}$Mg is partly because of smaller deformation than others; see Table 1 of Ref.~\cite{yos09a}. 
Another reason is the weak-binding effect, as discussed below. 
The experimental data indicate that the pairing in $^{36,38}$Mg would be 
weaker than the calculation because the measured $\mathcal{J}^{(1)}$ is 
slightly larger than the calculated one. 
Furthermore, the measurement shows a faster increase in $\mathcal{J}^{(1)}$ than the calculation 
for $^{38}$Mg; the calculation produces a stronger pairing correlation.

\begin{figure}[t]
\begin{center}
\includegraphics[scale=0.5]{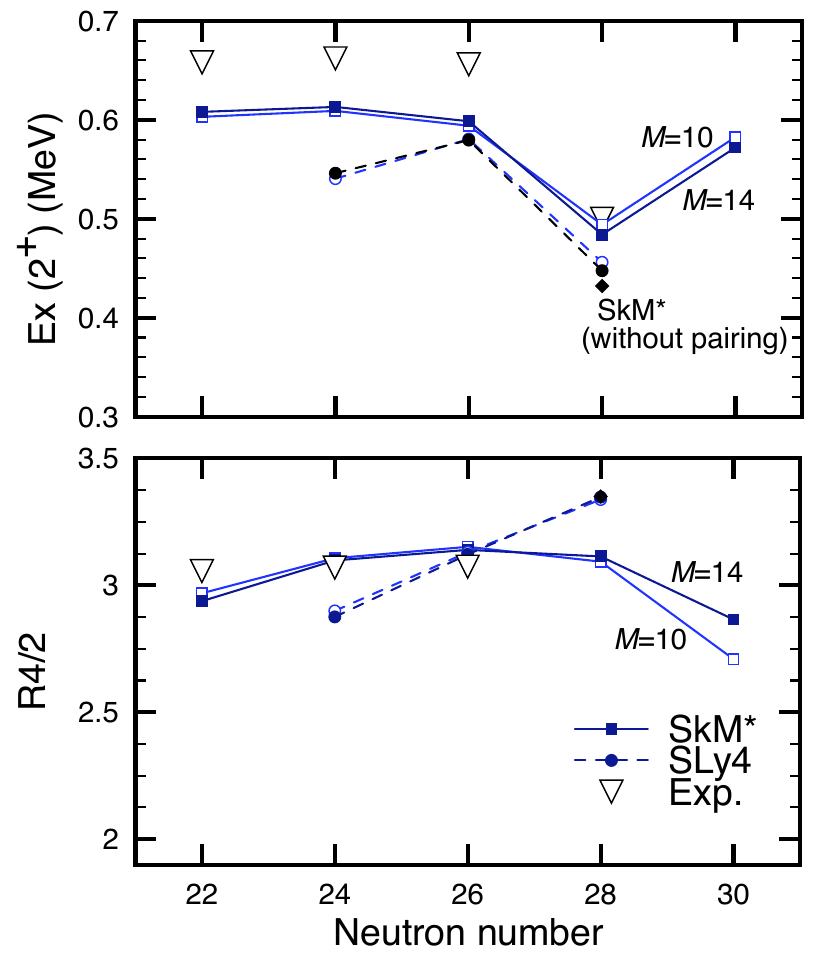}
\caption{\label{fig:Mg2} 
Calculated $E(2^+)$ (upper) and $R_{4/2}$ (lower) values with the SkM* (squares) and SLy4 (circles) functionals. 
The results obtained by using $M=10$ and 14 are depicted by open and filled symbols, respectively. 
The result without the pairing is shown by a diamond for SkM* at $N=28$. 
The experimental data are displayed by open triangles. 
}
\end{center}
\end{figure}

\begin{figure}[t]
\begin{center}
\includegraphics[scale=0.35]{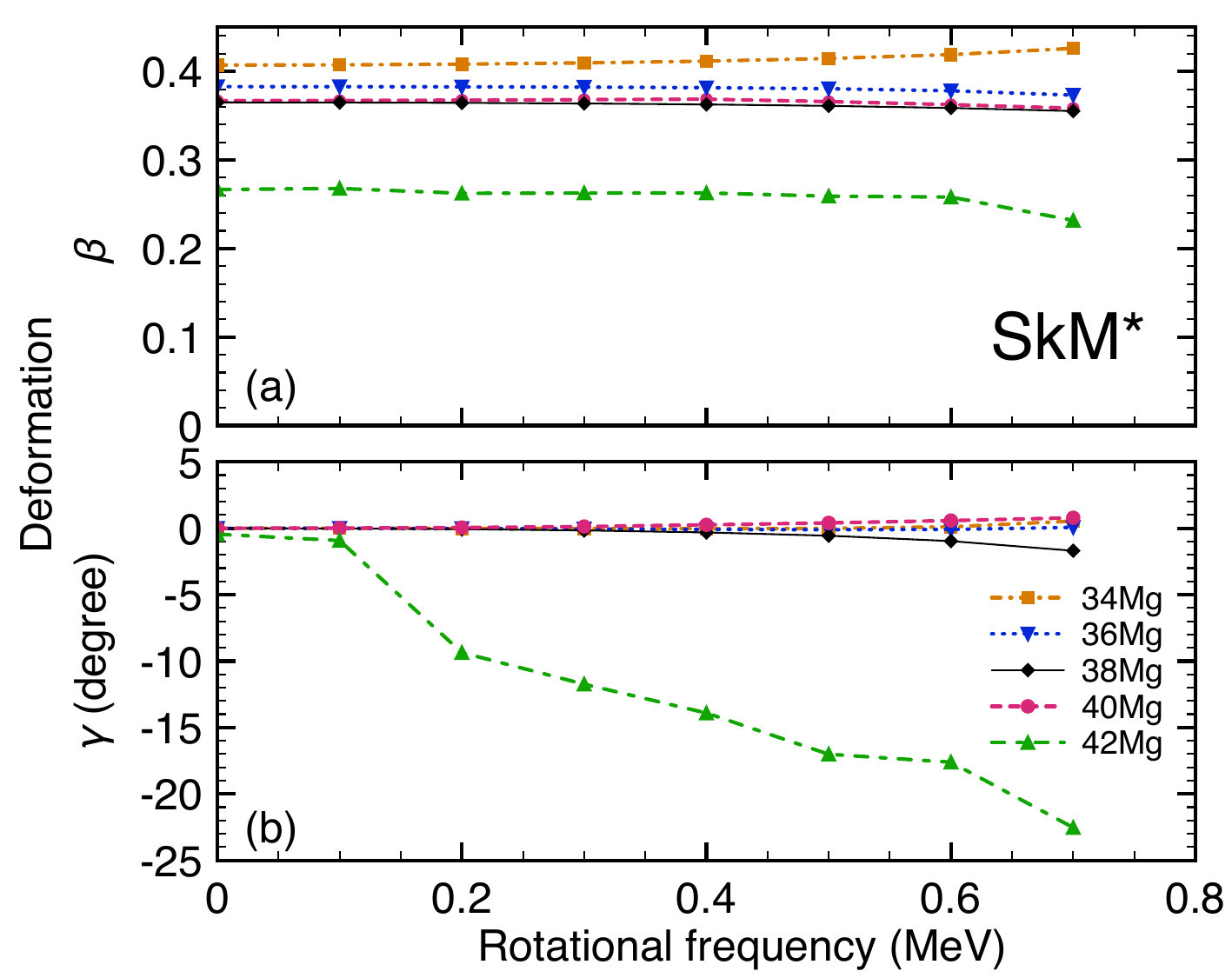}
\caption{\label{fig:Mg_def} 
Deformation parameters of protons in $^{34,36,38,40}$Mg.  
}
\end{center}
\end{figure}

The present calculation fails to describe the rotational motion in $^{40}$Mg: 
the calculated $\mathcal{J}^{(1)}$ is far below the measured value. 
One is then tempted to expect that the weak-binding effect shows up. 
In Fig.~\ref{fig:Mg}(b), I show the results obtained by varying the box size. 
With the increase in the box size, the calculated $\mathcal{J}^{(1)}$ increases. 
The calculation with $M=8$ does not show the convergence, 
while the difference between the results with $M=10$ and 12 becomes small.  
One cannot expect a further increase in $\mathcal{J}^{(1)}$ even enlarging the box size more. 
Notice that for the case of $^{34}$Mg, the results obtained with $M=8$ and 10 
are not very different, as shown in Fig.~\ref{fig:34Mg}. 
In this sense, the weak-binding effect appears in $^{40}$Mg; 
the weak binding of neutrons reduces the pairing correlation and enhances the moment of inertia. 
As an extreme case of weak pairing, I performed the calculation without the pairing, 
as displayed in Fig.~\ref{fig:Mg}(b). 
The resultant $\mathcal{J}^{(1)}$ is much larger than that obtained with pairing. 
As increasing the rotational frequency, the results with and without pairing become closer to each other. 
The observed $\mathcal{J}^{(1)}$ of 6.0 MeV$^{-1}$ is between these results 
and is relatively closer to the one obtained without pairing $\mathcal{J}^{(1)}=6.8$ MeV$^{-1}$.
As pointed out in Ref.~\cite{ter97}, one has an oblate minimum in $^{40}$Mg. 
Indeed, I found the oblate solution with $\beta=0.16, \gamma=60^\circ$ 
though this is located higher in energy by 0.83 MeV.
With this configuration, the calculated $\mathcal{J}^{(1)}$ is much lower: $1.9$ MeV$^{-1}$.

To describe the isotopic dependence of low-spin states in the Mg isotopes, 
a key is the isospin dependence of the pairing correlation. 
In Ref.~\cite{mar07}, it has been pointed out that 
the inclusion of the isospin dependence in the pairing EDF gives a 
nice reproduction of the pairing gaps in both stable and neutron-rich nuclei and in both symmetric nuclear matter and in neutron matter. 
Thus, I employ the optimal pairing EDF, the YSN functional~\cite{yam09} as above, 
in which the isovector density is introduced to describe pairing in nuclei in a wide mass region. 
Figure~\ref{fig:Mg2} shows the 
evaluated $E(2^+)$ and $R_{4/2}$ values. 
Here, the spin $I$ is evaluated as $J_z^2=I(I+1)$.

An almost-constant $E(2^+)$ and $R_{4/2}$ in $N=22\text{--}26$ is well described 
by using the SkM*$+$YSN functional. 
This model also describes well the decrease in energy at $N=28$ 
whereas this predicts the $R_{4/2}$ value keeps $\sim 3$.
When the pairing is discarded, $E(2^+)$ becomes much lower, and $R_{4/2}$ reaches $3.3$ as depicted by a diamond.
The results obtained by employing the SLy4$+$YSN functional are also shown 
in Fig.~\ref{fig:Mg2}. 
The SLy4$+$YSN predicts that neutrons and protons are both unpaired in $^{40}$Mg. 
Thus, this gives a similar result to the SkM* model without pairing. 
Whether the pairing of neutrons vanishes or not is discriminated by the $R_{4/2}$ value~\footnote{If the spin of the 
observed second excited state is $4^+$, the $R_{4/2}$ value is 2.34~\cite{cra19}.}.

When two neutrons are added, I find a further structural change.  
Comparing with $^{40}$Mg, the deformation gets weaker by about 33\%, and the triaxiality emerges as shown in Fig.~\ref{fig:Mg_def}, 
where the sign of $\gamma$ is defined in the convention of Ref.~\cite{nil95}. 
The triaxiality develops as increasing the spin. 
The appearance of the triaxiality in the $N=30$ isotones 
has also been discussed in Ref.~\cite{suz21}. 
The pairing for protons shows up, 
and that for neutrons increases.
Accordingly, $\mathcal{J}^{(1)}$ is reduced and thus $E(2^+)$ increases 
as shown in Fig.~\ref{fig:Mg2}. 
The present calculation with SkM*$+$YSN predicts 
that the irregularity in $E(2^+)$ appears only at $^{40}$Mg. 
Because of the developed triaxiality at finite $\omega_{\rm rot}$, the $R_{4/2}$ value 
deviates from that of the rigid rotor in $^{42}$Mg. 
It is noted that the SLy4 functional gives $\lambda^\nu >0$ for $^{42}$Mg.

A weak-binding effect is investigated by varying the box size 
and the role in the low-spin states is displayed in Fig.~\ref{fig:Mg2}. 
The results obtained by using $M=10$ are compared with those obtained by using $M=14$. 
I found that the calculated $\mathcal{J}^{(1)}$ at $\omega_{\rm rot}=0.1$ MeV changes by about 1.3\% when the box size is varied 
for $^{42}$Mg. 
Since the chemical potential of neutrons is not very shallow, that is $-1.15$ MeV for the case of SkM*, 
a role of the spatial extension of neutrons is not visible in the $E(2^+)$ value.  
However, the spatial expansion varies depending on $\omega_{\rm rot}$ 
as one finds that the $R_{4/2}$ value becomes smaller by 6\% in enlarging the box size from $M=10$ to 14.
Investigation of not only the $2^+$ state but higher-spin states in drip-line nuclei 
reveals unique roles of loosely bound neutrons. 
A triaxialy-deformed rotating nucleus is, for example, interesting future work to study~\cite{uza21}.

\section{Conclusion}

To summarize, 
I have developed a numerical framework for a mean-field description of yrast states 
in nuclei near the drip line in a nuclear EDF approach. 
To this end, I directly solved the coordinate-space cranked Skyrme-KSB equation in 3D mesh, 
with the continuum states being discretized in a box. 
The present framework reproduces the low-spin states and band crossing in a medium-heavy deformed nucleus.

The low-spin states in $^{34}$Mg are well described by using the density-dependent pairing interaction. 
With the increase in the neutron number, the calculation overestimates the pairing correlation, 
thus leading to the underestimation of the moments of inertia. 
Employing the optimal pairing-EDF constructed to describe neutron-rich nuclei, 
I have found that the appearance of the significantly low $I^\pi=2^+$ state in $^{40}$Mg 
is mainly due to the suppression of pairing. 
A systematic study of the $2^+$ state in neutron-rich nuclei thus provides a constraint on the global pairing EDF.
In $^{42}$Mg, the $2^+_1$ state appears higher in energy than in $^{40}$Mg, 
and the $R_{4/2}$ value decreases 
due to the structure change, where the triaxial deformation emerges.

\begin{acknowledgments} 
This work was supported by the JSPS KAKENHI (Grants No. JP19K03824 and No. JP19K03872). 
The numerical calculations were performed on the computing facilities  
at the Yukawa Institute for Theoretical Physics, Kyoto University, 
and at the Research Center for Nuclear Physics, Osaka University. 

\end{acknowledgments}

%

\end{document}